# Dynamic behaviour of Multilamellar Vesicles under Poiseuille flow

A. Pommella,[a] D. Donnarumma,[a] S. Caserta,[b,c,d*] and S. Guido[b,c,d]

a. *Laboratoire Charles Coulomb, Université de Montpellier & CNRS; Place Eugène Bataillon, 34095, Montpellier, France.*

b. *Dipartimento di Ingegneria Chimica dei Materiali e della Produzione Industriale (DICMAPI) Università di Napoli Federico II, P.le Tecchio, 80, 80125 Napoli, Italy.*

c. *CEINGE Advanced Biotechnologies, via G. Salvatore 486, 80145 Napoli, Italy.*

d. *Consorzio Interuniversitario Nazionale per la Scienza e Tecnologia dei Materiali (INSTM), UdR INSTM Napoli Federico II, P.le Tecchio, 80, 80125 Napoli, Italy.*

\* *Corresponding author: sergio.caserta@unina.it*

Surfactant solutions exhibit multilamellar surfactant vesicles (MLVs) under flow conditions and in concentration ranges which are found in a large number of industrial applications. MLVs are typically formed from a lamellar phase and play an important role in determining the rheological properties of surfactant solutions. Despite the wide literature on the collective dynamics of flowing MLVs, investigations on the flow behavior of single MLVs are scarce. In this work, we investigate a concentrated aqueous solution of linear alkylbenzene sulfonic acid (HLAS), characterized by MLVs dispersed in an isotropic micellar phase. Rheological tests show that the HLAS solution is a shear-thinning fluid with a power law index dependent on the shear rate. Pressure-driven shear flow of the HLAS solution in glass capillaries is investigated by high-speed video microscopy and image analysis. The so obtained velocity profiles provide evidence of a power-law fluid behaviour of the HLAS solution and images show a flow-focusing effect of the lamellar phase in the central core of the capillary. The flow behavior of individual MLVs shows analogies with that of unilamellar vesicles and emulsion droplets. Deformed MLVs exhibit typical shapes of unilamellar vesicles, such as parachute and bullet-like. Furthermore, MLV velocity follows the classical Hetsroni theory for droplets provided that the power law shear dependent viscosity of the HLAS solution is taken into account. The results of this work are relevant for the processing of surfactant-based systems in which the final properties depend on flow-induced morphology, such as cosmetic formulations and food products.

# 1. INTRODUCTION

The study of surfactant-based systems is relevant for both their wide scientific interest and broad industrial applications. Surfactant systems are ubiquitous in everyday life: most detergents, lubricants, cosmetics and many pharmaceutical and food products are surfactant-based. In particular, the alkyl benzene sulfonate group is one of the most commonly used moiety of anionic detergents in industrial applications.[1, 2] In this class, the linear alkyl benzene sulfonic acid (HLAS) represents the most widely used anionic surfactant in industrial products, with physical and mechanical properties strongly dependent on the manufacturing process.[3, 4]

In aqueous solution, surfactant molecules, such as HLAS, due to their amphiphilic nature, self-assemble into a range of well-defined structures depending on temperature, concentration and their chemical nature.[5, 6] These structures range from micellar solutions to multilamellar phases. In wide concentration ranges, multilamellar domains can fold rearranging in more or less spherical multilamellar structures made of onion-like stacked bilayers separated by solvent, known as surfactant MultiLamellar Vesicles (MLVs).[7-10]

The behaviour of surfactant MLVs has been studied under different flow conditions in order to characterize their properties for several applications, such as bio-membrane models[11] or carriers for drug delivery.[12, 13] In literature, many studies describe the flow behaviour of MLVs under shear flow. MLVs collective behaviour has been investigated with the aim of finding scaling relationships[14] between their size and an external flow field by several techniques, including small-angle neutron scattering[15] and polarized light microscopy.[16] We recently reported an experimental investigation on the dynamic behaviour of single MLVs under simple shear flow.[17, 18] In this flow field, MLVs dynamics presents analogies with unilamellar vesicles and emulsion droplets. MLVs deform at constant volume under flow with a rich dynamic behaviour (i.e. tumbling, breathing and tank-treading), which is also observed in unilamellar vesicles.[19, 20] Furthermore, we found that MLVs can be characterized by an effective interfacial tension $\sigma_{eff}$ whose value can be determined by analysing their deformation in analogy with emulsion droplets.[21]

While most studies on MLVs have been addressed to simple shear flow, a detailed analysis concerning the behaviour of MLVs (and more in general of multiphase aqueous surfactant solutions) under pressure-driven (or Poiseuille) flow in micro-channels is still missing. Among the other questions, one may ask whether the analogies between the dynamic behaviour of MLVs and unilamellar vesicles or droplets are also found under capillary flow. As briefly reviewed in the following, the flow behaviour of unilamellar vesicles and emulsion droplets in micro-channels have been extensively studied and their dynamics is well known.

Unilamellar vesicles and capsules are often considered as a model system for blood flow in microcirculation[22, 23] and for red blood cell (RBC) membrane deformation.[24-26] This is one of the reasons motivating the investigation of the dynamics of single unilamellar vesicles and capsules in Poiseuille flow.[27-30] Recent studies have highlighted that the unique mechanical properties of the lipid bilayer membrane of unilamellar vesicles, such as fluidity, incompressibility, and resistance to bending, give rise to a number of fascinating non-equilibrium features of vesicle and RBC micro hydrodynamics, namely, parachute and bullet shapes.[24, 31] Therefore, it would be interesting to evaluate whether MLVs could also be considered as a model system to mimic blood flow.[32, 33]

The behaviour of emulsion droplets in confined Poiseuille flow has been widely investigated as well.[34] Briefly, the difference between droplet and suspending fluid velocity has been quantified by measuring the droplet velocity $U$, normalized with respect to the average value $V_m$ of the undisturbed parabolic velocity profile of the external fluid far away from the droplet. In case of Newtonian fluid components, droplet velocity depends on the two classical non-dimensional parameters, first introduced by Taylor[35] to describe flow-induced deformation of single droplets. These parameters are the viscosity ratio $\lambda$ between the viscosity $\eta_d$ of the droplet phase and the viscosity $\eta$ of the continuous phase, and the capillary number $Ca$, which is defined as the ratio between the viscous and interfacial stresses acting on the droplet. In tube flow, the capillary number is given by the expression $Ca = \eta V_m / \sigma$, where $\sigma$ is the interfacial tension between the two immiscible fluids and $V_m$ is the undisturbed average velocity of the continuous phase. In the case of confined flow, a further geometrical parameter is introduced[36] as the ratio $k = D / D_c$ between droplet ($D$) and channel ($D_c$) diameter. For non-Newtonian fluids, some additional non-dimensional parameters are needed to take into account viscoelastic contributions, and can be expressed as the ratio between fluid relaxation time and emulsion time.[37] From a theoretical point of view, Hetsroni et al.[38] investigated the case of small undeformed and unconfined droplets ($k \ll 1$) moving along the axis of a circular cross-section channel. They calculated the non-dimensional droplet velocity $U / V_m$, showing that it lies in between the maximum and the average velocities of the undisturbed flow. From the experimental and numerical side, the problem has been tackled[39-41] by investigating the shape of a deformed droplet in a range of values of the non-dimensional parameters $k$, $\lambda$ and $Ca$. They confirmed the Hetsroni theory finding that, at small $k$ values, the ratios $U / V_m$ of droplets deformed at different $Ca$ superimpose on the same master curve showing that the only relevant parameter is the confinement ratio $k$. They also investigated the droplet deformation and velocity in conditions far from Hetsroni theory ($k \approx 1$) finding that the ratio $U / V_m$ does not follow anymore the Hetsroni theory and new trends appear depending on the Capillary numbers. Indeed, a uniform film forms between the droplet and the wall for sufficiently large droplets.[41, 42] The film thickness is determined by the balance between viscous and capillary forces and depends only on $Ca$ and $\lambda$.

In this work, we report an investigation on the behaviour of an aqueous HLAS solution under Poiseuille flow by direct flow visualization techniques. After a rheological characterization in oscillatory regime and in continuous flow of the HLAS solution, we will show how the microstructure of the sample changes when a Poiseuille flow is imposed. In particular, we will focus on the dynamic behaviour of individual multilamellar vesicles analysing their shape and their velocity under confined flow. To our knowledge, this is the first experimental investigation on MLV dynamics under Poiseuille flow. Understanding MLVs flow behaviour in pressure-driven flow, in particular under confined conditions, would be quite relevant for microfluidic applications[43] either to produce materials[44] or to mimic biological conditions[24] and flow at the microscale.[45, 46]

## 2. EXPERIMENTAL

### 2.1. Materials

The sample examined in this work is an aqueous solution of linear alkyl benzene sulfonic acid (HLAS), a commercial anionic surfactant constituted by a benzene ring linking a sulfonic acid group (hydrophilic head) and a linear alkyl group (hydrophobic tail).[47] The fluid is a commercial sample of industrial grade and is used without any further purification. The phase diagram of the HLAS-water system[48] can be divided in three different regions depending on HLAS concentration and temperature. At the temperature of 25° C, the phase diagram shows an isotropic micellar ($L_1$) phase at HLAS concentrations lower than 28% wt. At higher HLAS concentrations (> 67% wt.) the surfactant molecules form planar lamellae and the system is constituted by a lamellar ($L_\alpha$) phase. Between these two concentration values, both $L_1$ and $L_\alpha$ phases coexist and $L_\alpha$ domains can fold forming MLVs.[9, 48, 49] A 30% wt. solution of HLAS is used in our experiments at 25° C. This HLAS concentration falls within the intermediate region and the sample shows MLVs ranging from a few to hundreds of microns as observed by optical microscopy.[17] MLVs were observed under static conditions in the HLAS solution used in this work, possibly due to the stirring flow applied during preparation.

### 2.2. Methods

#### 2.2.1. Rheology.
The rheological measurements of the sample are carried out by a stress-controlled rotational MCR 301 rheometer (Anton Paar Instruments). A titanium double-gap measuring system (DG 26.7, inner cup diameter 24.267 mm, inner bob diameter 24.666 mm, outer bob diameter 26.663 mm, outer cup diameter 27.053 mm, and bob height: 40.000 mm) is used with 1.8 ml of solution. Both continuous and oscillatory flow tests are performed at 25°C as reported in the results section.

#### 2.2.2. Rheo-optical apparatus.
The analysis of MLVs flow behaviour under capillary flow is performed using a rheo-optical apparatus. It consists of a flow chamber made up of silica micro-capillaries (Polymicro Technologies) that is placed under the field of view of an optical microscope (Axioscope FS, Zeiss). Micro-capillaries with different internal diameter $D_c$ (106, 180, and 318 μm) are tested and a syringe pump (Harvard 11 Plus) is used to feed the HLAS solution in the micro-channels using a feeding tube of diameter $D_f$ = 1 mm. The syringe pump and the micro-capillaries are coupled by microfluidics connections (Upchurch, IDEX). Images of MLVs flowing inside the micro-capillaries are acquired through a high-speed video camera (Phantom 640) and analysed off-line. Observations are done in bright field using long working distance objectives (10x and 20x). The whole apparatus is placed on an anti-vibration table in a temperature-controlled room (T = 25±1°C).

#### 2.2.3. Image analysis.
Fluid morphology under flow is quantified by image analysis techniques using a commercial software (Image Pro Plus 7). The images are acquired using a high-speed video camera and stored on the computer hard drive in digital format. Average diameter $D$, velocity $U$ and shape of vesicles under flow are determined by identifying manually the MLV outline (long dashed line) as shown in Figure 1. Velocity profiles in the channel are determined by tracking at different radial positions the trajectories of MLVs with diameter smaller than 1/10 of the capillary radius, which are used as tracers. The height $\delta$ of the central section of the capillary, where MLVs migrate under flow, is measured as shown in Figure 1 and normalized with respect to the capillary diameter $D_c$. The edges of the central section (short dashed lines) are chosen by selecting the position of the flowing MLVs closest to the capillary walls for each experimental condition.

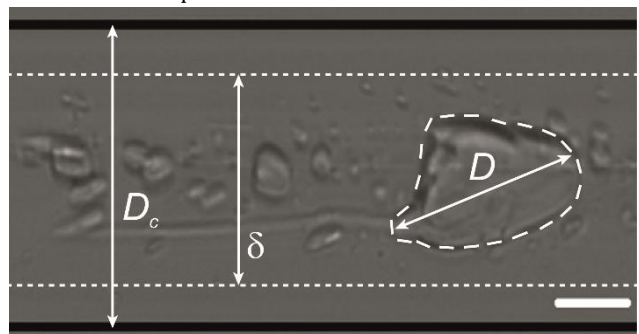

Figure 1 - The image shows MLVs of the HLAS solution (30% wt.) in capillary flow (dimensional bar: 40 μm). The main geometrical dimensions considered for data analysis are illustrated in the image: average diameter of the deformed vesicle $D$, capillary inner diameter $D_c$, diameter $\delta$ of the central section of the capillary where the vesicles are observed to migrate under flow.

## 3. RESULTS AND DISCUSSION

The experimental investigation of the 30% wt. HLAS aqueous solution is carried out by probing both the macroscopic and microscopic behaviour of the sample. In the following, we start from the macroscopic rheological behaviour of the sample and its dependence on the internal microstructure. We will then focus on the velocity profiles and the flow-induced segregation of the different phases of the HLAS sample. In the final part of the results section, the sample microscopic behaviour will be investigated by looking at the dynamics of single MLVs under capillary flow.

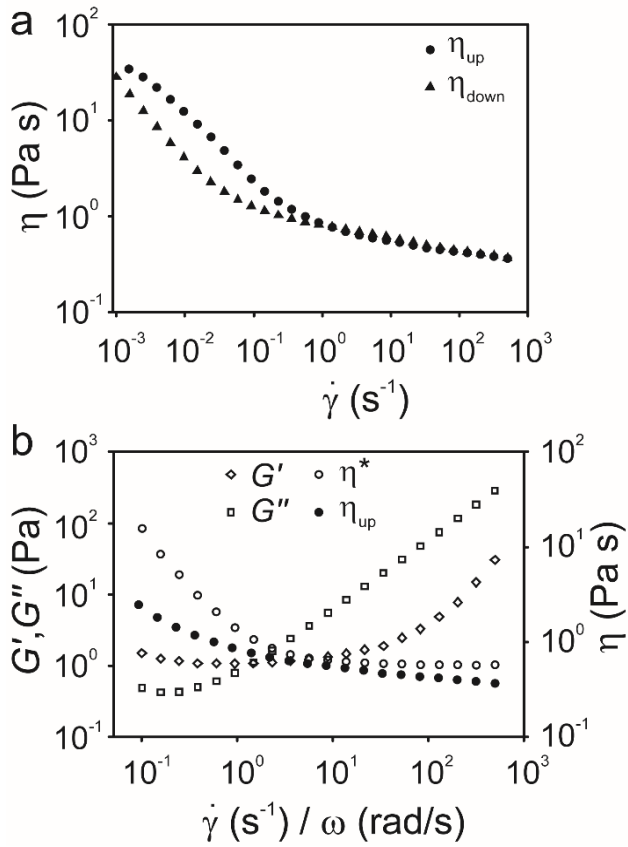

Figure 2 - Rheological characterization of 30% wt. HLAS-water solution. a) The flow curve shows a shear-thinning behaviour of the viscosity with a hysteresis loop. The shear-thinning regime is characterized by the presence of two different slopes corresponding to a power low index $n \approx 0.3$ and $n = 0.87$ for low and high shear rates, respectively. b) In the oscillatory tests a gel-like behaviour is visible in the low frequencies range ($\omega$ < 1 rad/s) with the elastic modulus $G'$ higher than the viscous modulus $G''$. At higher frequencies a fluid-like behaviour is observed with $G''$ higher than $G'$. Complex viscosity data do not superimpose with steady shear viscosity measurement.

## 3.1. Rheological results.

The rheological characterization of the HLAS sample is shown in Figure 2. In Figure 2a, the flow curve shows a shear-thinning behaviour and the presence of hysteresis when the viscosity is measured going first from low to high shear rates and then from high to low shear rates. The two sets of data (referred to as $\eta_{up}$ and $\eta_{down}$ respectively in Figure 2a) superimpose at high shear rates (> 1 s$^{-1}$), but a significant discrepancy is observed at low shear rates (< 1 s$^{-1}$). Such hysteresis, which is a non-Newtonian feature, is typically found when a weak microstructure is present in the sample at rest, such as in weak gels. Flow-induced break-up of the weak microstructure is consistent with the finding that $\eta_{up}$ > $\eta_{down}$ at low shear rates. Once the microstructure is broken in the "Up" run, the sample exhibits a lower viscosity in the subsequent "Down" run since it has not been allowed to equilibrate between the two runs and to recreate the original microstructure at rest. The shear rate where $\eta_{up}$ and $\eta_{down}$ start overlapping (ca 1 s$^{-1}$) can be taken as the threshold value for complete breakup of the weak microstructure. In our sample, this behaviour is due to the arrangement of the lamellar L$_\alpha$ phase that forms a polydomain microstructure made by a network of smectic liquid crystalline monodomains (planar or spherical, i.e. MLVs) with random orientation.[50] This type of microstructures was already observed through polarized microscopy and transmission electron microscopy in several non-ionic surfactants water solutions[51] and we previously[18] showed the same polycrystalline microstructure for the HLAS surfactant as well.

Interestingly, an internal microstructure made by dispersed L$_\alpha$ domains has been reported at a concentration even lower than 5%.[52] In the flow curve of Figure 2a, it is also interesting to notice the presence of two different shear-thinning in the range of shear rates investigated. The transition between the two slopes is coincident with the threshold value of approximately 1 s$^{-1}$ whereupon the $\eta_{up}$ and $\eta_{down}$ curves overlap. By fitting the shear-thinning behaviour with a power law curve ($\boldsymbol{\eta = m\dot{\gamma}^{n-1}}$), two values of the power law index $n$ can be estimated for the two slopes. In the low shear rate range ($\boldsymbol{\dot{\gamma}}$ < 0.15 s$^{-1}$), data fitting gives a power law index $n \approx 0.3$ ($n = 0.36$ and $0.26$ for "Up" and "Down", respectively), while in the second range ($\boldsymbol{\dot{\gamma}}$ > 0.15 s$^{-1}$) a value of $n = 0.87$ is obtained from both the "Up" and "Down" measurements. The presence of two slopes in the viscosity curve can be attributed to different dynamics during the rearranging of the internal microstructure depending on the value of the imposed shear rate, in agreement with the hysteresis behaviour. Therefore, the first slope might be due to the deformation and breakage of the lamellar domains. This happens because an external flow tends to orient the lamellar domains along the direction of the flow but the final degree of orientation depends on the intensity of the flow applied (i.e. the value of the applied shear rate).[53]

In Figure 2b the elastic and viscous moduli and the complex viscosity (open symbols) are plotted as a function of frequency $\omega$, in the range $10^{-1}$ – $10^3$ rad/s. Data at lower frequencies result to be outside the range of linear viscoelasticity and are not reported. The frequency tests confirm the presence of a weak microstructure in our HLAS sample due to the lamellar domains. Indeed, in the low frequency range ($\omega$ < 1 rad/s) a gel-like behaviour is visible, with both moduli showing a parallel horizontal trend. The elastic modulus $G'$ is higher than the viscous one $G''$, indicating a dominant elastic behaviour of the fluid. At $\omega$ > 1 rad/s the viscous modulus $G''$ predominates and grows linearly with slope $\approx 1$, while the elastic modulus $G'$ remains almost constant up to $\omega \approx 10$ rad/s and then grows with a slope $\approx 1.4$. A $G'$ - $G''$ crossover occurs at frequency $\omega \approx 1.5$ rad/s. The complex viscosity $\eta^*$ (open circles in Figure 2.b) exhibits a shear-thinning behaviour similar to the one shown by the shear viscosity $\eta_{up}$ (filled circles in Figure 2.b), also plotted for comparison. The complex viscosity shows a region with a relevant shear-thinning ($n \approx 0.1$) in the low frequencies range (up to $\omega \cong 1$ rad/s), while higher frequencies data result in an almost constant value of $\eta^*$ ($n \approx 1$). However, despite the similar qualitative trend shown by the complex and shear viscosity, it should be noticed that the Cox-Merz equation does not hold for this system, i.e., continuous and oscillatory viscosity data do not superimpose quantitatively at the same frequency/shear rate.

## 3.2. HLAS microstructure under Poiseuille flow.

In Figure 3a, b, c we show three experimental velocity profiles measured in the micro-capillaries tested in our experiments. In Figure 3d the shear rate profiles are shown as function of the capillary radius for the three profiles. Experimental data of the profiles are obtained by measuring the velocity of isolated MLVs (points in Figure 3) smaller than 1/10 of the capillary radius which are considered as tracers. The uncertainty of the experimental data is rather lower as shown in Figure 3a, b, c by the size of the error bars which is about the same as the size of the data points. The experimental data are compared to the expected velocity profiles according to the rheological bulk measurements (dashed lines in Figure 3a, b, c). In order to obtain these profiles, we combined the equation of the stress balance in a circular tube,[54] $\tau = -\Delta P r / 2L$, and the constitutive equation $\tau(r) = \eta(r)\dot{\gamma}(r)$, obtaining:

$$\dot{\gamma}(r) = -\frac{dv}{dr} = -\frac{\Delta P}{L}\frac{1}{\eta(r)}\frac{r}{2} \qquad \text{(Eq. 1)}$$

where $\Delta P / L$ is the pressure difference between the capillary inlet and the outlet per unit of capillary length. Eq. 1 was integrated numerically over the radius with no-slipping boundary conditions. We estimated the dependence of viscosity on the radius $r$ by coupling the function $\dot{\gamma} = f(r)$, obtained from a power-law fitting of the experimental velocity profiles, with the flow curve $\eta_{down}$ shown in Figure 2a. The observation that the velocity has a power law dependency with the radius is equivalent to the hypothesis that the velocity profile is the one expected from a power-law fluid, described by Eq. 2:

$$v = v_{max}\left(1 - \left(\frac{r}{R_c}\right)^{\frac{n+1}{n}}\right) \qquad \text{(Eq. 2)}$$

where $v_{max} = \frac{n}{n+1}\left(\frac{\Delta P}{2mL}\right)^{\frac{1}{n}}R_c^{\frac{n+1}{n}}$, and $m$ and $n$ are the two parameters of the power-law fluid introduced in the previous section. The assumption that the relation $\dot{\gamma} = f(r)$ obtained from the fit of the velocity profile can be used to integrate Eq. 1 is based on the microviscometric method,[55] that was experimentally validated for RBCs flowing in 50 μm diameter micro-channels.

The comparison between the experimental and expected (according to rheology) velocity profiles (dashed lines) shows a good agreement only for the largest micro-capillary (Figure 3c) with larger discrepancies for the other cases (Figure 3a, b). On the other hand, it can be observed that the fit of the experimental data using the velocity profile of a power-law fluid (Eq. 2), reported as continuous line in Figure 3a, b, c, is in good agreement with the behaviour of the HLAS solution in micro-capillary for all the conditions investigated.

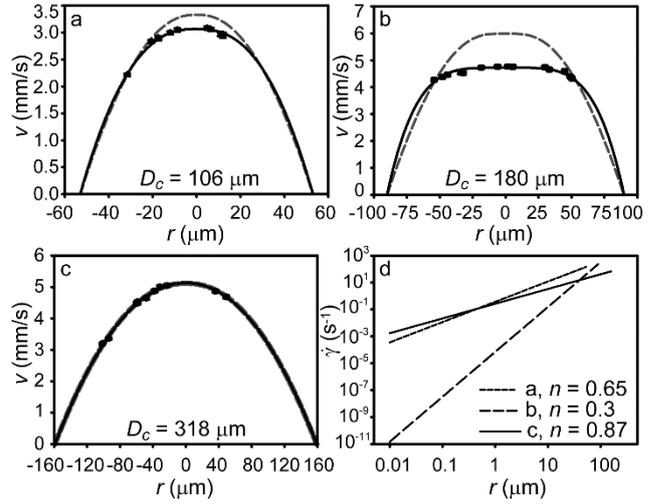

Figure 3 - a, b, c) The velocity profiles of HLAS samples for each micro-capillary are shown. Black circles are the experimental measurements, dashed lines are the theoretical velocity profiles obtained from the rheological measurements, solid lines are the fit of the profiles using the velocity profile equation of a power-law fluid (Eq. 2). Flow rate $Q$, maximum velocity $V_{max}$ and wall shear rate $\dot{\gamma}_w$ are: a) Q = 54 μL/h, $V_{max}$ = 3.06 mm/s, $\dot{\gamma}_w$ = 145 s$^{-1}$; b) Q = 296 μL/h, $V_{max}$ = 4.72 mm/s, $\dot{\gamma}_w$ = 227 s$^{-1}$; c) Q = 750 μL/h, $V_{max}$ = 5.13 mm/s, $\dot{\gamma}_w$ = 67 s$^{-1}$. d) Shear rate profiles of the three velocity profiles, shown in a log-log scale. The different slope of each profile reflects the different power-law index $n$.

The fit of Eq. 2 shows three different values of the power-law index $n$ for the three data sets reported in Figure 3, and the range 0.3-0.87, shown in Figure 3c, represents the overall range explored in our experiments. As shown by the dashed lines in Figure 3, the viscosity data coming from rheology fail to explain all the experimental velocity profiles apart those characterized by a value of $n$ close to 0.87. There are several possible explanations for this discrepancy. One reason could be a flow-induced rearrangement of the micellar phase due to the extensional flow at the entrance of the micro-capillaries. This effect would be less important for large micro-capillary diameter given the smaller value of the contraction ratio $D_c/D_f$, where $D_f$ is the diameter of the tube feeding the capillary. This would explain the fact that, for the larger micro-capillary, the velocity profile is well described by a power-law fluid profile with a value of $n$ that is in agreement with the viscosity data obtained from rheology. Indeed, looking at the shear rate profile shown in Figure 3d, we observe that, in the case of Figure 3c, most of the sample is subject to values of shear rate falling in the range 0.1 s$^{-1}$ – 67 s$^{-1}$, with the exception of a channel core of size 1 μm. Within this range of shear rate the viscosity shows a power-law shear thinning behaviour (Figure 2a) with an index $n$ of 0.87 in agreement with the index value obtained from the power-law fit of the velocity profile. On the contrary, with different values of the imposed flow rate and capillary diameter, the initial sample microstructure could be more significantly affected, leading to velocity profiles that can be still described by a power-law fluid equation, but with a smaller value of $n$ that is no longer in agreement with the viscosity data of Figure 2a. Another possible explanation of the discrepancy between velocity profiles and viscosity data is the onset of wall slip at some critical value of wall shear rate in the micro-capillaries. This effect could be less significant in the rheological tests since the double gap device is made of stainless steel as opposed to silica. In any event, the elucidation of the discrepancy is beyond the scope of this work.

Entrance effects are also likely to have an impact on the lamellar phase of the HLAS solution, as exemplified in Figure 4, where the MLVs are observed to concentrate in the central core of the micro-capillaries. In Figure 4 images of flowing MLVs are shown for each micro-capillary diameter. The values of maximum fluid velocity $V_{max}$ and power-law index $n$ is also indicated in each image. In Figure 4 it is shown that MLVs (and more generally the $L_\alpha$ phase) accumulate around the central axis in a region of height $\delta$, defined in Figure 1. It can be observed that the height $\delta$ becomes smaller and smaller by increasing the fluid velocity $V_{max}$. We investigated this flow-focusing effect in a quantitative way by measuring $\delta$ as function of the mean velocity $V_m \times D_c/D_f$. It is worth remembering that the maximum velocity is related to the mean velocity $V_m$ of the velocity profile by the relation $\frac{V_{max}}{V_m} = \frac{1+3n}{1+n}$ for a power-law fluid. A lower limit of the volume fraction of the central region is given by the volumetric concentration of the $L_\alpha$ phase in the HLAS solution, which can be estimated as (30-28)/(67-30)~0,05 based on the concentrations of the micellar (28%) and lamellar (67%) phases. This limiting value was approached for all the micro-capillary diameters used in the experiments.

A possible explanation of this separation is that, at the entrance of the micro-capillary, the fluid streamlines approach each other towards the centre of the micro-capillary. Hence, the curvature of the flow lines can induce radial inward migration of deformable objects, such as vescicles,[56] thus concentrating the $L_\alpha$ phase at the centre. This effect can be taken as proportional to the mean velocity $V_m$ and to the contraction ratio $D_c/D_f$, where $D_f$, the diameter of the tube feeding the capillary, is equal to 1 mm in all the experiments. This scaling is assessed in Figure 5 where the volume occupied by the lamellar phase in the central region ($V_{L\alpha} = \frac{\pi}{4}\delta^2 L$), normalized to the capillary volume V ($V = \frac{\pi}{4}D_C^2 L$), is plotted versus the quantity $V_m \times D_c/D_f$. Data in Figure 5 collapse on the same master curve, thus supporting the proposed scaling. It is worth mentioning that this entrance effect does not imply that images were acquired too close to the micro-capillary entrance. Indeed, as shown in Figure 3, the velocity profiles are fully developed without any influence of the entrance on the flow streamlines. Once MLVs are in the centre of the micro-capillary, they do not migrate back towards the walls probably because of their deformability that interacts with the nonlinearity of the Poiseuille flow, resulting in a cross-streamline migration of vesicles toward the centre of the channel. [30, 57] This effect has been shown for soft object such as droplets,[58, 59] unilamellar vesicles[60] and capsules[29] and is due to the presence of an inhomogeneous velocity gradient in the transverse direction of the flow.

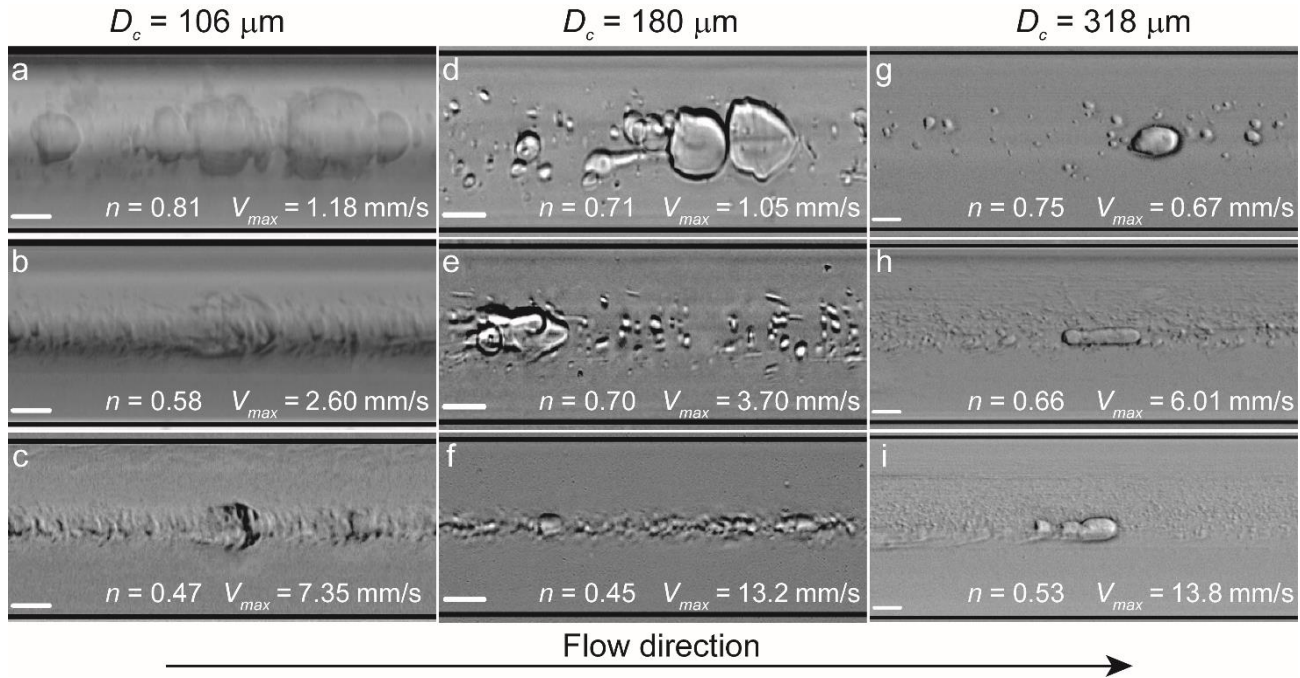

Figure 4 - Images of MLVs flowing in the micro-capillaries at different maximum fluid velocities $V_{max}$ and power-law index $n$ for each micro-capillary diameter $D_c$. The figure shows that the height of the section $\delta$ where the MLVs are grouped is a decreasing function of the velocity $V_{max}$ within each micro-capillary. Dimensional bars are 40 μm.

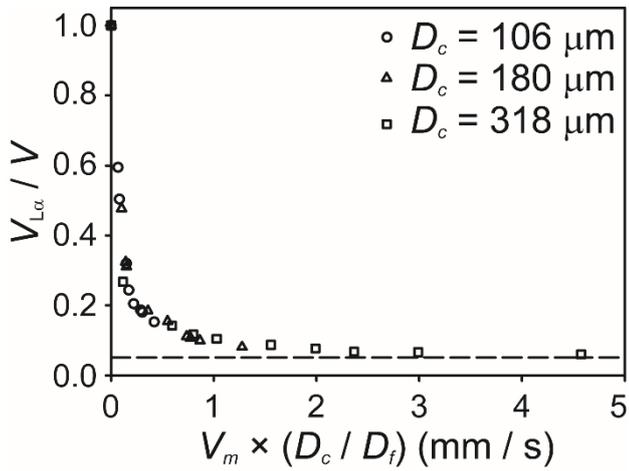

Figure 5 - Scaling of the volume occupied by MLVs ($V_{L\alpha}$) in the micro-capillary, normalized to the volume of the micro-capillary ($V$), versus the quantity $V_m \times (D_c / D_f)$ for different fluid mean velocities and capillary diameters. The dashed line represents the volumetric concentration of the $L_\alpha$ phase in the HLAS solution and corresponds to the plateau value approached at high values of $V_m \times (D_c / D_f)$.

### 3.3. MLV dynamics under Poiseuille flow.

In Figure 6, the shape of MLVs flowing along the centreline is shown as a function of the capillary number $Ca = \eta V_m / \sigma_{eff}$ and the confinement degree $k = D / D_c$. It can be observed that some MLVs do not display an axial-symmetric shape, typical of soft objects flowing along the centreline under Poiseuille flow. This is likely due to the presence of defects in the structure of MLVs (e.g., images c, d, g, h in Figure 6); these defects have also been reported in a previous paper.[18] It is worth mentioning that the presence of these internal defects does not allow MLVs to assume an axial-symmetric shape even at high value of the confinement degree (Figure6 c, f, i). In any event, MLV shapes are apparently more axial-symmetric at increasing values of $Ca$. In the top left images, the capillary number and the confinement degree are small and a parachute-like or bullet-like shape is found. These two shapes are typical of both droplets[41] and unilamellar vesicles[30] and they are described here for the first time for MLVs. It should be noticed that the typical concavity at the rear found in droplets and in unilamellar vesicles is not found in MLVs. Strongly deformed shapes are not observed even at the highest values of $Ca$ and $k$ investigated. This could be due to the lower deformability of the MLV structure, which is made of a stack of several bilayers, as compared to the single bilayer structure of unilamellar vesicles.[18] The formation of filaments coming out of MLVs suggests a break-up mechanism similar to droplet tip-streaming (image g in Figure 6).

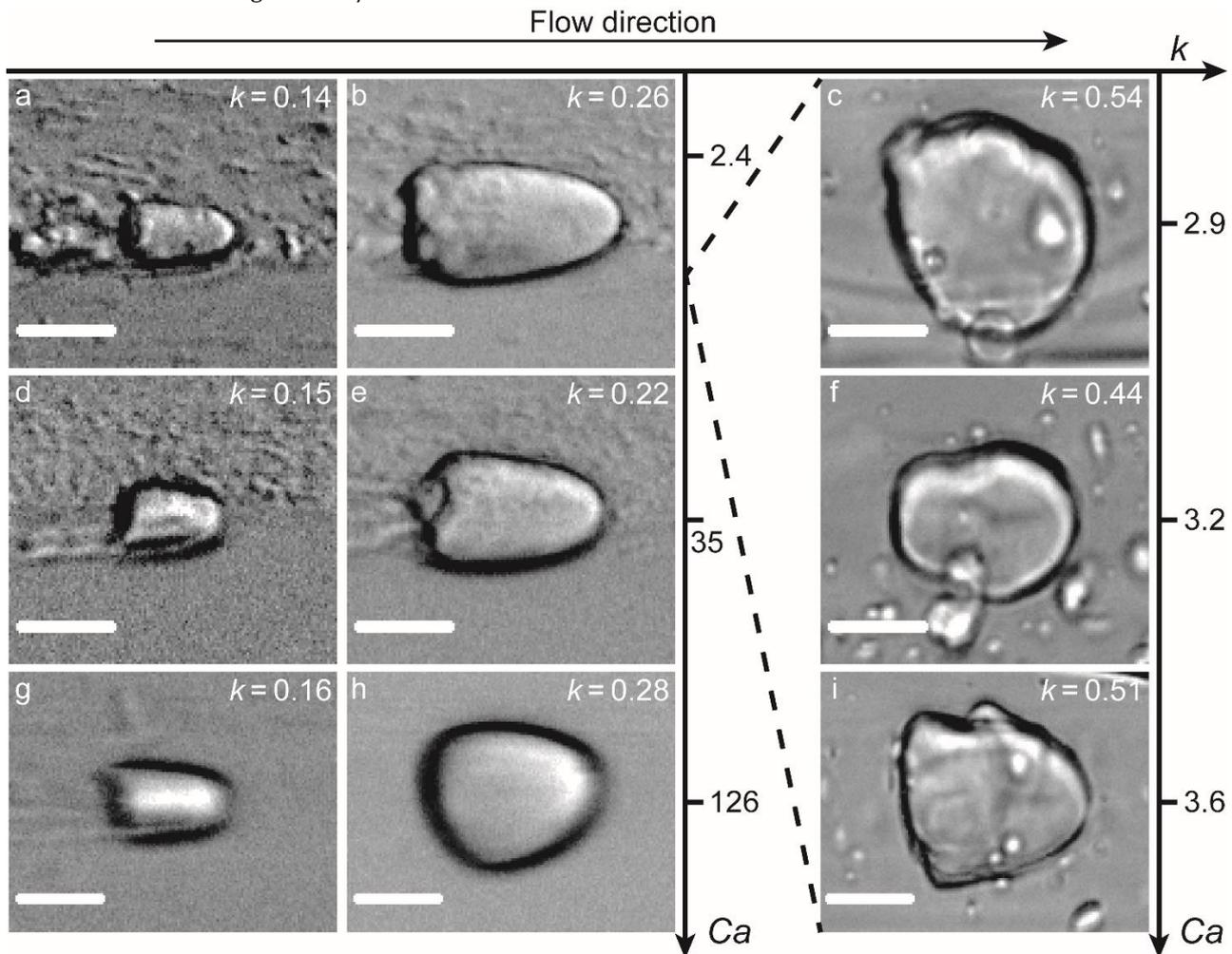

Figure 6 - Shapes of deformed MLVs under Poiseuille flow at different capillary number $Ca$ and confinement degree $k$. Dimensional bars are 40 μm.

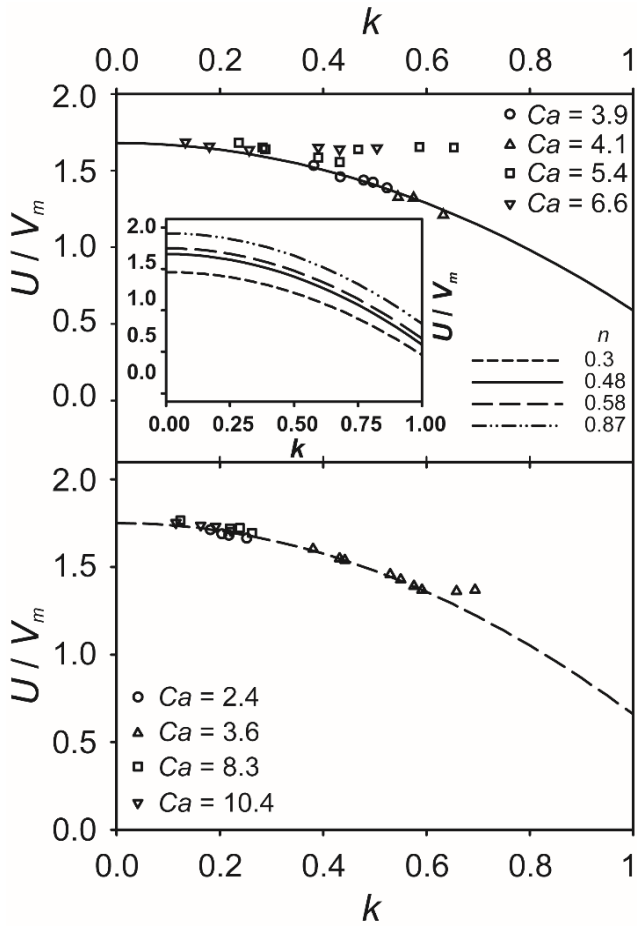

Figure 7 - Scaling of the MLVs velocity $U$ normalized to the fluid mean velocity $V_m$ with the confinement degree $k$ parametric in the capillary number $Ca$. The lines are fitting of the data according to Hetsroni theory modified for a power law fluid. a) Data corresponding to a power law index $n = 0.48$ are shown. The inset shows the range of variation of the theoretical curves in the range of $n$ between 0.3 and 0.87. b) Data refer to a power law index $n = 0.58$.

A quantitative investigation of the dynamic behaviour of MLVs under Poiseuille flow is carried out by analysing the MLV velocity $U$ as shown in Figure 7. The velocity $U$, normalized to the average flow velocity in the micro-capillary $V_m$, is plotted as a function of the confinement parameter $k$ for different values of the capillary number $Ca = \eta V_m / \sigma_{eff}$, where we use the value of $\sigma_{eff} = 0.1$ mN/m reported in a previous paper.[17] This is the classical scaling used for droplets and it appears to be valid for MLVs as well. In the case of unilamellar vesicles, a distinction needs to be made. Generally, two capillary numbers are defined for each of the two elastic constants governing the membrane elasticity, the bending and stretching modulus ($\kappa_b$ and $\kappa_s$ respectively).[61] If viscous stresses are low enough, the bending modulus dominates the dynamics and $Ca = \eta V_m R^2 / \kappa_b$. On the other hand, when viscous stresses are sufficiently large, the stretching modulus becomes dominant and the capillary number is usually defined as in the droplet case with $\kappa_s$ instead of $\sigma_{eff}$.[27] In our MLVs experiments, we can infer from the observed scaling in Figure 7 that a bending modulus is not likely to play a significant role in the MLVs dynamics. Regarding the stretching modulus, it is typically of the order of hundreds of mN/m in unilamellar vesicles. Using this value we obtain $Ca \approx 10^{-3}$ which is not consistent with the significant MLVs deformations observed in the experiments. On the other hand, more realistic values of $Ca \approx 1$ are calculated with $\sigma_{eff}$. Therefore, we believe that $\sigma_{eff}$ is the elastic constant governing MLVs dynamics in our flow conditions because viscous stresses are of the same order of magnitude of the membrane elasticity given by the effective interfacial tension arising from MLVs multilamellar structure.[14]

As shown in Figure 7 MLVs velocity follows the Hetsroni theory[38] developed for emulsion droplets in confined Poiseuille flow. In particular, the droplet velocity $U$, normalized to the mean velocity of the fluid in the circular capillary $V_m$, is well represented by:

$$\frac{U}{V_m} = p - \frac{4\lambda}{3\lambda+2}k^2 + O(k^3) \qquad \text{(Eq. 3)}$$

where $p$ is the ratio between the maximum and the mean velocity in the capillary, $k$ is confinement degree, and $\lambda$ is the ratio between the droplet and matrix viscosity. The term $O(k^3)$ is a higher order infinitesimal and is typically neglected. Eq. 3 states that the velocity of the droplet decreases as the square power of $k$, while the viscosity ratio $\lambda$ determines how the confinement degree affects droplet velocity. Indeed, for $\lambda \to 0$ the velocity is independent from the confinement degree while for $\lambda > 1$ we have the square power dependence.

The term $p$ in Eq. 3 is the ratio between the maximum and the mean velocity in the capillary. In the case of Newtonian fluids, this term is equal to 2 while, in the case of a power law fluid, it becomes $p = (1+3n) / (1+n)$. We have previously observed that the HLAS solution flowing in micro-capillaries can be modelled as a power law fluid with a power law index $n$ ranging from 0.3 to 0.87 (Figure 3) depending on the range of shear rate.

In Figure 7a we show the experimental data characterized by a power law index $n = 0.48$ obtained from the fitting of the velocity profiles. Considering the corresponding velocity ratio $p = 1.65$, Eq. 3 was used to fit the experimental data using $\lambda$ as the only fitting parameter. A different set of data, relative to a value of $n = 0.58$ ($p = 1.75$) is plotted in Figure 7b. The value of $\lambda$ obtained from the fitting of the Hetsroni predictions from the two set of data in Figure 7 is about 2 and should be considered as an average quantity due to the shear-thinning behaviour of the HLAS solution viscosity. This value is in agreement with the value of $\lambda$ obtained from the deformation of MLVs under shear flow.[17] It is worth mentioning that a different value of the parameter $n$ determines only a different value of the velocities ratio $p$ in Eq. 3 without any influence on the trend of the curve. To appreciate this aspect, in the inset of Figure 7a, the two curves relative to $n = 0.48$ and $n = 0.58$ are compared together with the case of $n = 0.3$ and $n = 0.87$, reported as a reference to show the global range of variation of $n$ for our samples.

In Figure 7 the normalized velocity $U / V_m$ collapses on the same master curve for different $Ca$ and $k$, in agreement with the Hetsroni scaling developed for emulsion droplets. The agreement is good for low values of confinement degree, while at higher $k$ there are deviations from the theoretical predictions suggesting some levelling off to a plateau. These deviations are expected, being the Hetsroni theory valid for small undeformed droplets, and they are also observed for droplets. In particular, Lac et Sherwood[41] found that a uniform film forms between the droplet and the micro-capillary wall for sufficiently large droplets. The film thickness does not depend on the confinement degree $k$ but only on $Ca$ and $\lambda$ due to the balance between viscous and capillary forces acting on the droplets. This film allows droplets to move at a constant velocity above a certain confinement degree explaining the plateau of the normalized velocity $U / V_m$. In their work, they showed that $U / V_m$ deviates from the Hetsroni theory around a minimum value of $k \approx 0.5$ for $\lambda = 1$ and $Ca = 0.5$. In our experiments on MLVs we have higher values of $Ca$ and we observe deviations from $k = 0.6$ at $Ca = 3.6$ (Figure 7b) to $k = 0.4$ at $Ca = 5.4$ (Figure 7a). In the case of unilamellar vesicles, it has been found that they follow the Hetsroni theory too when bending energies are negligible.[27] However, the presence of the plateau at high confinement ratios is not observed probably due to the very small value of $Ca$ involved ($Ca \approx 10^{-6}$). This observation seems to support our idea that the effective interfacial tension $\sigma_{eff}$ is the main elastic constant making the MLVs dynamics more similar to droplets than unilamellar vesicles in our experimental conditions.

## 4. CONCLUSIONS

In this paper, an experimental investigation of a HLAS aqueous solution under pressure-driven flow in micro-capillaries is carried out by probing both the macroscopic and microscopic behaviour of the sample. The HLAS solution is set to a concentration of 30% wt. where only a lamellar and a micellar phases are present. The lamellar phase generates MLVs whose dynamics determine the solution behaviour under flow. We start from the macroscopic rheological behaviour of the sample and its dependence on the internal microstructure. Secondly, we focus on the velocity profiles and the flow-induced segregation of the different phases of the HLAS sample. In the final part, the sample microscopic behaviour is investigated by looking at the dynamics of single MLVs under capillary flow. The rheological characterization of the HLAS solution shows that it can be described as a power law fluid characterized by two regions with different power law index and by the presence of hysteresis, likely due to flow-induced breakage of solution microstructure. The velocity profiles show that the solution behaves as a power-law fluid under capillary flow. The rheological characterization agrees with the velocity profiles only for the largest micro-capillary. The discrepancy observed with the small micro-capillaries is mainly attributed to a structure rearrangement of the solution due to the contraction ratio of the micro-capillary. The microstructural rearrangement is observed under flow with a segregation of the lamellar and micellar phases. The lamellar phase forming MLVs accumulates in the central part of the micro-capillaries while the micellar MLV-free phase is found closer to the walls. A scaling of this segregation with the product between the fluid mean velocity and the micro-capillary diameter is proposed. This scaling is explained in terms of fluid streamlines approaching to the centre of the micro-capillary at the entrance of the micro-capillary.

The investigation of the dynamics of single MLVs shows several analogies with the behaviour of unilamellar vesicles and emulsion droplets under Poiseuille flow. MLVs are observed to deform assuming a parachute-like or bullet-like shape depending on the confinement degree and the capillary number; a shape diagram is reported showing how these two parameters influence MLV shape. The analysis of MLV velocity under Poiseuille flow provides evidence that the effective interfacial tension is the main elastic parameter of MLV membranes in analogies with droplets. The velocity is found to follow the Hetsroni theory developed for droplets. By adapting the theory to a power law fluid, a scaling of the MLV velocity with the capillary number and the confinement degree is presented.

The results of this work are relevant for the processing of surfactant-based systems, where the final product properties are strongly dependent on the flow-induced morphology.

## CONFLICTS OF INTEREST

There are no conflicts of interest to declare.

## ACKNOWLEDGEMENTS

This work has been done under the umbrella of COST Action MP1106 Smart and Green interfaces for biomedical and industrial applications and of COST Action MP1305 Flowing Matter. We thank Dr. Antonio Perazzo for helpful discussions, Valeria Mazzola for her support to the experiments during her master thesis, and Vincenzo Guida from P&G for helpful discussions and for providing HLAS samples.

## REFERENCES

1. K. Masters, *Spray Drying Handbook*, Longman Scientific & Technical, 1991.

2. J. I. Kroschwitz and A. Seidel, *Kirk-Othmer Encyclopedia of Chemical Technology*, Wiley, 2006.
3. J. L. Berna, C. Bengoechea, A. Moreno and R. S. Rounds, *Journal of Surfactants and Detergents*, 2000, **3**, 353-359.
4. H. W. Stache, *Anionic Surfactants: Organic Chemistry*, Taylor & Francis, 1995.
5. V. Guida, *Advances in Colloid and Interface Science*, 2010, **161**, 77-88.
6. E. Kaler, A. Murthy, B. Rodriguez and J. Zasadzinski, *Science*, 1989, **245**, 1371-1374.
7. M. S. Liaw, M. R. Mackley, J. Bridgwater, G. D. Moggridge and A. E. Bayly, *AIChE Journal*, 2003, **49**, 2966-2973.
8. S. A. McKeown, M. R. Mackley and G. D. Moggridge, *Chemical Engineering Research and Design*, 2003, **81**, 649-664.
9. C. Richards, G. J. T. Tiddy and S. Casey, *Langmuir*, 2007, **23**, 467-474.
10. A. Perazzo, V. Preziosi and S. Guido, *Advances in Colloid and Interface Science*, 2015, **222**, 581-599.
11. S. J. Singer and G. L. Nicolson, *Science*, 1972, **175**, 720-731.
12. S. Kessler, R. Finken and U. Seifert, *Journal of Fluid Mechanics*, 2008, **605**, 207–226.
13. M. N. Holme, I. A. Fedotenko, D. Abegg, J. Althaus, L. Babel, F. Favarger, R. Reiter, R. Tanasescu, P.-L. Zaffalon, A. Ziegler, B. Muller, T. Saxer and A. Zumbuehl, *Nat Nano*, 2012, **7**, 536-543.
14. S. Fujii and W. Richtering, *The European Physical Journal E*, 2006, **19**, 139-148.
15. S. Koschoreck, S. Fujii, P. Lindner and W. Richtering, *Rheologica Acta*, 2009, **48**, 231-240.
16. Z. Yuan, S. Dong, W. Liu and J. Hao, *Langmuir*, 2009, **25**, 8974-8981.
17. A. Pommella, S. Caserta, V. Guida and S. Guido, *Physical Review Letters*, 2012, **108**, 138301.
18. A. Pommella, S. Caserta and S. Guido, *Soft Matter*, 2013, **9**, 7545-7552.
19. M. Guedda, *Physical Review E*, 2014, **89**, 012703.
20. C. Misbah, *Physical Review Letters*, 2006, **96**, 028104.
21. S. Caserta, S. Reynaud, M. Simeone and S. Guido, *Journal of Rheology*, 2007, **51**, 761-774.
22. G. Tomaiuolo, M. Barra, V. Preziosi, A. Cassinese, B. Rotoli and S. Guido, *Lab on a Chip*, 2011, **11**, 449-454.
23. P. M. Vlahovska, D. Barthes-Biesel and C. Misbah, *Comptes Rendus Physique*, 2013, **14**, 451-458.
24. G. Tomaiuolo, M. Simeone, V. Martinelli, B. Rotoli and S. Guido, *Soft Matter*, 2009, **5**, 3736-3740.
25. L. Lanotte, J. Mauer, S. Mendez, D. A. Fedosov, J.-M. Fromental, V. Claveria, F. Nicoud, G. Gompper and M. Abkarian, *Proceedings of the National Academy of Sciences*, 2016, **113**, 13289-13294.
26. D. Barthès-Biesel, *Annual Review of Fluid Mechanics*, 2016, **48**, 25-52.
27. V. Vitkova, M. Mader and T. Podgorski, *EPL (Europhysics Letters)*, 2004, **68**, 398.
28. G. Coupier, A. Farutin, C. Minetti, T. Podgorski and C. Misbah, *Physical Review Letters*, 2012, **108**, 178106.
29. S. Nix, Y. Imai and T. Ishikawa, *Journal of Biomechanics*, 2016, **49**, 2249-2254.
30. G. Danker, P. M. Vlahovska and C. Misbah, *Physical Review Letters*, 2009, **102**, 148102.
31. D. Abreu, M. Levant, V. Steinberg and U. Seifert, *Advances in Colloid and Interface Science*, 2014, **208**, 129-141.
32. T. Savin, M. M. Bandi and L. Mahadevan, *Soft Matter*, 2016, **12**, 562-573.
33. J. M. Barakat, A. P. Spann and E. S. Shaqfeh, presented in part at 69th Annual Meeting of the APS Division of Fluid Dynamics, Portland (OR), November, 2016.
34. S. Guido and V. Preziosi, *Advances in Colloid and Interface Science*, 2010, **161**, 89-101.
35. G. I. Taylor, *Proceedings of the Royal Society of London. Series A*, 1934, **146**, 501-523.
36. M. Shapira and S. Haber, *International Journal of Multiphase Flow*, 1990, **16**, 305-321.
37. S. Guido and F. Greco, *Rheology Reviews*, 2004, 99-142.
38. G. Hetsroni, S. Haber and E. Wacholder, *Journal of Fluid Mechanics*, 1970, **41**, 689-705.
39. B. P. Ho and L. G. Leal, *Journal of Fluid Mechanics*, 1975, **71**, 361-383.
40. M. J. Martinez and K. S. Udell, *Journal of Fluid Mechanics*, 1990, **210**, 565-591.
41. E. Lac and J. D. Sherwood, *Journal of Fluid Mechanics*, 2009, **640**, 27-54.
42. S. R. Hodges, O. E. Jensen and J. M. Rallison, *Journal of Fluid Mechanics*, 2004, **501**, 279-301.
43. G. M. Whitesides, *Nature*, 2006, **442**, 368-373.
44. S. L. Anna, *Annual Review of Fluid Mechanics*, 2016, **48**, 285-309.
45. A. Jahn, W. N. Vreeland, M. Gaitan and L. E. Locascio, *Journal of the American Chemical Society*, 2004, **126**, 2674-2675.
46. V. Preziosi, A. Perazzo, G. Tomaiuolo, V. Pipich, D. Danino, L. Paduano and S. Guido, *Soft Matter*, 2017, Accepted Manuscript.
47. W. M. Linfield, *Anionic Surfactants*, M. Dekker, 1976.
48. J. A. Stewart, A. Saiani, A. Bayly and G. J. T. Tiddy, *Colloids and Surfaces A: Physicochemical and Engineering Aspects*, 2009, **338**, 155-161.
49. A. Sein, J. B. F. N. Engberts, E. van der Linden and J. C. van de Pas, *Langmuir*, 1996, **12**, 2913-2923.
50. J. Muñoz and M. C. Alfaro, *Grasas y Aceites*, 2000, **51**, 6-25.
51. R. Heusch and F. Kopp, *Berichte der Bunsengesellschaft für physikalische Chemie*, 1987, **91**, 806-811.
52. J. Szymański, A. Wilk, R. Hołyst, G. Roberts, K. Sinclair and A. Kowalski, *Journal of Non-Newtonian Fluid Mechanics*, 2008, **148**, 134-140.
53. M. G. Berni, C. J. Lawrence and D. Machin, *Advances in Colloid and Interface Science*, 2002, **98**, 217-243.
54. R. B. Bird, W. E. Stewart and E. N. Lightfoot, *Transport Phenomena*, Wiley, 2007.
55. D. S. Long, M. L. Smith, A. R. Pries, K. Ley and E. R. Damiano, *Proceedings of the National*

*Academy of Sciences of the United States of America*, 2004, **101**, 10060-10065.
56. G. Ghigliotti, A. Rahimian, G. Biros and C. Misbah, *Physical Review Letters*, 2011, **106**, 028101.
57. B. Kaoui, G. H. Ristow, I. Cantat, C. Misbah and W. Zimmermann, *Physical Review E*, 2008, **77**, 021903.
58. P. C. H. Chan and L. G. Leal, *Journal of Fluid Mechanics*, 1979, **92**, 131-170.
59. V. Preziosi, G. Tomaiuolo, M. Fenizia, S. Caserta and S. Guido, *Journal of Rheology*, 2016, **60**, 419-432.
60. A. Farutin, T. Piasecki, A. M. Slowicka, C. Misbah, E. Wajnryb and M. L. Ekiel-Jezewska, *Soft Matter*, 2016, **12**, 7307-7323.
61. A. Pommella, N. J. Brooks, J. M. Seddon and V. Garbin, *Scientific Reports*, 2015, **5**, 13163.